# Tuning of gain layer doping concentration and Carbon implantation effect on deep gain layer

S. M. Mazza*, C. Gee, Y. Zhao, R. Padilla, E. Ryan, N. Tournebise, B. Darby
F. McKinney-Martinez, H. F.-W. Sadrozinski, A. Seiden, B. Schumm
*SCIPP, Univ. of California Santa Cruz, 1156 High Street, Santa Cruz CA 95064, USA*

V. Cindro, G. Kramberger, I. Mandić, M. Mikuž, M. Zavrtanik
*Jožef Stefan institute and Department of Physics, University of Ljubljana, Jamova cesta 39, 1000 Ljubljana, Slovenia*

R. Arcidiacono[1, 3], N. Cartiglia[1], M. Ferrero[1], M. Mandurrino[1], V. Sola[1, 2], A. Staiano[1]
[1]*INFN, Italy* [2]*Dipartimento di fisica, Universita' di Torino, Via Pietro Giuria 1, 10125 Torino (TO), Italy*
[3]*Universita' del Piemonte Orientale, Largo Donegani 1 Novara (NO), Italy*

M. Boscardin[1, 2], G.F. Della Betta[2, 3], F. Ficorella[1, 2], L. Pancheri[2, 3], G. Paternoster[1, 2]
[1]*Fondazione Bruno Kessler, Via Sommardive 18, 38123 Povo (TN), Italy* [2]*TIFPA-INFN, Via Sommardive 14, 38123 Povo (TN), Italy* [3]*Dipartimento di fisica, Universita' di Trento, Italy Via Sommardive 14, 38123 Povo (TN), Italy*

*Abstract*– Next generation Low Gain Avalanche Diodes (LGAD) produced by Hamamatsu photonics (HPK) and Fondazione Bruno Kessler (FBK) were tested before and after irradiation with ~1MeV neutrons at the JSI facility in Ljubljana. Sensors were irradiated to a maximum 1-MeV equivalent fluence of $2.5E15\ N_{eq}/cm^2$. The sensors analysed in this paper are an improvement after the lessons learned from previous FBK and HPK productions that were already reported in precedent papers. The gain layer of HPK sensors was fine-tuned to optimize the performance before and after irradiation. FBK sensors instead combined the benefit of Carbon infusion and deep gain layer to further the radiation hardness of the sensors and reduced the bulk thickness to enhance the timing resolution. The sensor performance was measured in charge collection studies using β-particles from a 90Sr source and in capacitance-voltage scans (C-V) to determine the bias to deplete the gain layer. The collected charge and the timing resolution were measured as a function of bias voltage at -30C. Finally a correlation is shown between the bias voltage to deplete the gain layer and the bias voltage needed to reach a certain amount of gain in the sensor. HPK sensors showed a better performance before irradiation while maintaining the radiation hardness of the previous production. FBK sensors showed exceptional radiation hardness allowing a collected charge up to 10 fC and a time resolution of 40 ps at the maximum fluence.

## 1. Introduction

Low Gain Avalanche Detectors (LGADs) are thin (20 to 60 μm) n-on-p silicon sensors with modest internal gain (typically 5 to 50) and exceptional time resolution (17 ps to 50 ps) [1-3]. LGADs were first developed by the Centro Nacional de Microelectrónica (CNM) Barcelona, in part as a RD50 Common Project [4]. The internal gain is due to a highly doped p+ region (called multiplication or gain layer) just below the n-type implants of the electrodes. The multiplication layer is up to a few microns thick, while the rest of the active area is referred to as the bulk. Thanks to their extraordinary properties LGADs establish a new paradigm for space-time particle tracking [5].

The first application of LGADs are planned for the High Luminosity LHC (HL-LHC [6]), where the extreme pileup conditions will lower the efficiency for tracking and vertexing of the inner tracking detector in the region close to the beam-pipe. Therefore, to maintain the performance, LGAD based timing layers are foreseen in the forward region of both the ATLAS and the CMS experiments. The two projects are called respectively the High Granularity Timing Detector (HGTD) [7] and the End-cap Timing Layer (ETL) [8]. At HL-LHC, LGADs would be of moderate segmentation (1.3 mm x 1.3 mm) and will have to face challenging radiation requirements, with fluences up to few $1E15\ N_{eq}/cm^2$ and doses up to few MGy.

* Corresponding author: simazza@ucsc.edu, telephone (831) 459 1293, FAX (831) 459 5777

LGADs from several vendors have been tested extensively during the last few years. Previous studies on LGAD sensors from different vendors are reported in [9-14]. In all cited cases, both the timing resolution and the gain deteriorate with radiation damage due to the acceptor removal mechanism [15, 16], which reduces the effective doping concentration in the gain layer. The performance loss from radiation damage can be reduced by implementing a deep gain layer or Carbon infusion in the sensor design [9, 10, 13]. In this paper the improvement in the new FBK and HPK productions after the lesson learned from previous prototypes runs will be shown.

Sensors shown in this study were irradiated with ~1 MeV neutrons at the Triga reactor in Lubjiana [17] to a maximum equivalent fluence of 2.5E15 $N_{eq}/cm^2$. The neutron energy spectrum and flux are well known. The fluence is quoted in 1 MeV equivalent neutrons per $cm^2$ ($N_{eq}/cm^2$) by using Non Ionizing Energy Loss (NIEL) scaling. After irradiation, the devices were annealed for 80 min at 60 C. Afterward the devices were kept in cold storage at -20 C to reduce further annealing. The irradiation fluence uncertainty is at most of the order of 10% as shown in [13].

## 2. Overview of tested sensors

The sensors in this study were produced by HPK (Japan) and FBK (Italy), Table 1 summarizes the characteristic of both the old and the new productions. In the new HPK production the doping concentration in the gain layer was reduced by 2% for each split, this is reflected in an increase of the breakdown voltage with the splits of 20-30V steps. All HPK sensors, old and new production, have a deep gain layer. In the new FBK production the normal active layer was decreased to 45 μm to improve the time resolution and the deep gain layer was implemented, furthermore the carbon implantation level was optimized. Several other wafers were produced by FBK (more details can be found in [18]) but this study is focused only on the three ones included in Table 1.

| Manufacturer | Type | Nominal thickness | VBD (-30C) | Carbon dose | Gain layer |
|---|---|---|---|---|---|
| HPK (old) | HPK-3.2 | 50 | 70 | 0 | Deep |
| FBK (old) | FBK3+C | 55 | 150 | 1 | Shallow |
| HPK | Split-1 | 50 | 90 | 0 | Deep |
| HPK | Split-2 | 50 | 110 | 0 | Deep |
| HPK | Split-3 | 50 | 150 | 0 | Deep |
| HPK | Split-4 | 50 | 170 | 0 | Deep |
| FBK | FBK3.2 W7 | 55 | 250 | 1 | Shallow |
| FBK | FBK3.2 W14 | 45 | 250 | 1 | Deep |
| FBK | FBK3.2W19 | 45 | - | 0.6 | Deep |

**Table 1:** Summary of new and old HPK and FBK productions

## 3. Experimental setup

The current over voltage (IV) and capacitance over voltage (CV) of the sensors is taken on a manual probe station with needles with a Keithley 2657A HV power supply and an Agilent E4980A capacimeter. The CV of the sensor is done at 10 kHz for non-irradiated sensors and at 1 kHz for irradiated sensors with a sinusoidal probe of 200 mV. The HV is applied from the back of the sensor through the chuck and the needles are put to ground. For the CV measurement the pad is read through the capacimeter and the guard ring (GR) is put directly to ground.

The charge collection experimental relies on a 90Sr beta-source, with a setup described in detail in [9-13]. The tested LGAD, defined as device under test (DUT), is mounted on a fast analog electronic board (up to 2 GHz bandwidth) digitized by a GHz bandwidth digital scope (Keysight DSO254A). The trigger, which acts as a time reference, is also mounted on a fast electronic board, and it is provided by a second HPK LGAD with time resolution of 17ps. The area of the pulse is evaluated for the DUT, subtracting the subsequent undershoot, then it is divided by the transimpedance of the amplifier system (4700 Ω) to calculate the collected charge. For the calculation of the time resolution a Constant Fraction Discriminator (CFD) of 50% is used to evaluate the time of arrival for the DUT and a CFD of 20% for the time of arrival of the trigger. The details of the analysis and the CFD method are reported in detail in [9-13].

## 4. Capacitance over voltage of irradiated and non-irradiated sensors (place holder plots)

From the $1/C^2$ measurement the $V_{GL}$ (voltage to deplete the gain layer) is extracted with the method explained in [10]. A $V_{GL}$ vs fluence distribution is then built for each type of sensor. The resulting plot is shown in Figure 1. The radiation damage on LGADs can be modelled with:

1. $N_A(\phi) = g_{eff}\phi + N_A(\phi=0)e^{-c\phi}$

Where the term $g_{eff}\phi$ is the acceptor creation mechanism by the creation of deep traps. The term $N_A(\phi=0)e^{-c\phi}$ instead is the initial acceptor removal mechanism that causes the reduction of doping concentration in the gain layer. The $N_A(\phi=0)e^{-c\phi}$ term is fitted to the distribution in Figure 1, and the C-factor (acceptor removal constant) is extracted for each detector type.

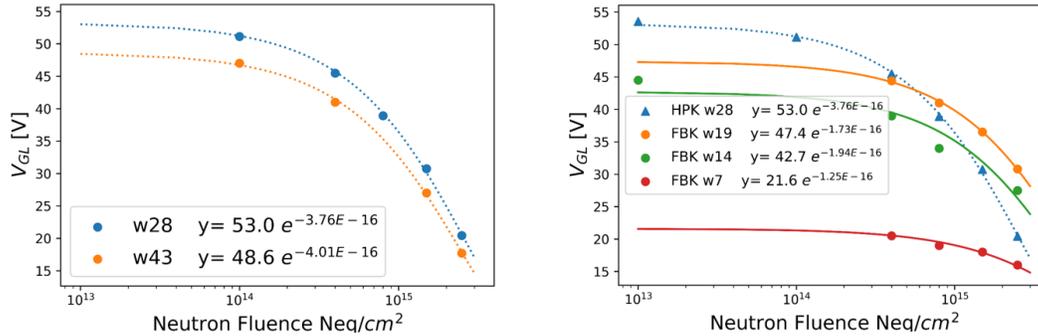

**Figure 1:** $V_{GL}$ vs fluence distribution for (Left) HPK split 1 and split 4 (Right) HPK split 1 and FBK W7, W14, W19. The fitted c-factor is shown in the legend.

## 5. Collected charge of irradiated and non-irradiated sensors

The performance of HPK sensors from split 1 and split 4 is shown in Figure 2, the collected charge (Left) of the two splits has a roughly constant separation in voltage. Before irradiation split 1 have a higher collected charge than split 4, this difference is reduced with radiation damage. The time resolution (Right) for split 4 is better before irradiation since the applied voltage to the bulk region is higher allowing for the saturation of hole velocity. However, after irradiation, both splits have the same performance reach in time resolution.
3

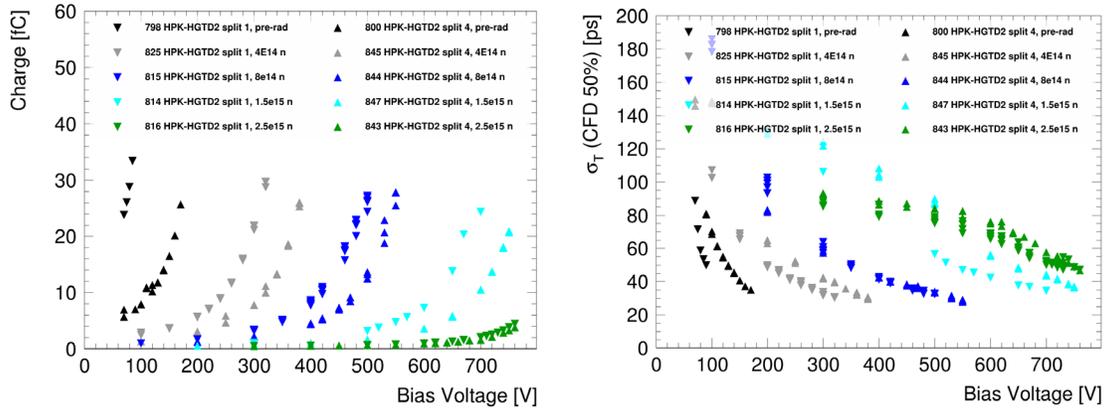

**Figure 2.** Collected charge (Left) and time resolution (Right) of HPK sensors split 1 and split 4 at all irradiation fluences.

The collected charge for all studied FBK wafer past and current production is shown in Figure 3 (Left). The gradual performance gain can be seen from one type to the next: the worst perfomance is of FBK3noC that is without Carbon, adding Carbon the performance is greatly improved (FBK3+C, FBK UFSD3.2 W7) and with a reduced sensor thickness (FBK UFSD3.2 W14) the same performance can be achieved by having a deep gain layer. Finally by optimizing the Carbon level and increasing the doping concentration of the deep gain layer the best performance can be achieved with FBK UFSD3.2 W19. In Figure 3 (Right) the time resolution at CFD 50% is shown for FBK W7 and W19 at all irradiation fluences, W19 have an improved time resolution thanks to the reduced bulk thickness.

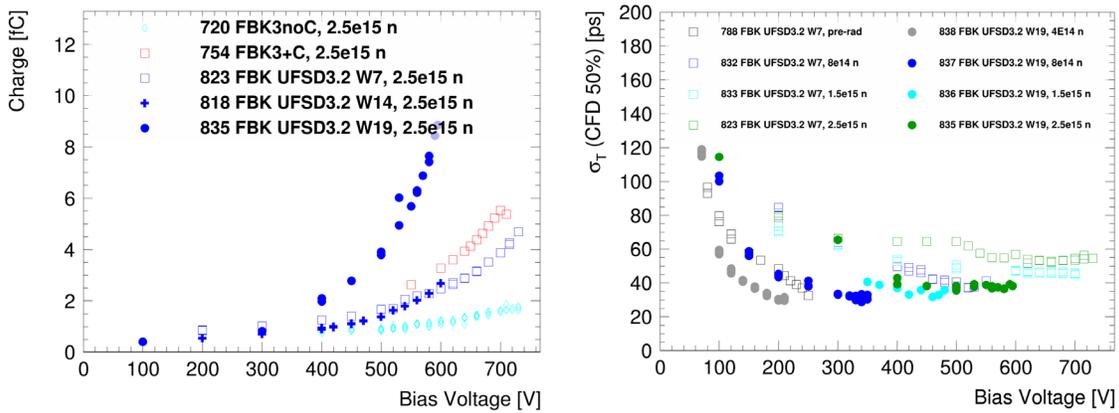

**Figure 3.** Collected charge (Left) for all types of tested FBK sensors at the fluence of 2.5E15 $N_{eq}/cm^2$ and time resolution (Right) of FBK UFSD3.2 sensor W7 and W19 at all irradiation fluences.

## 6. Correlation of CV and collected charge

The $V_{GL}$ value from the $1/C^2$ distribution is correlated with the bias voltage needed to achieve a collected charge of 4 fC ($V_{CC=4}$) in Figure 4. The correlation is linear as expected from previous studies, showing consistency across all the measurements.

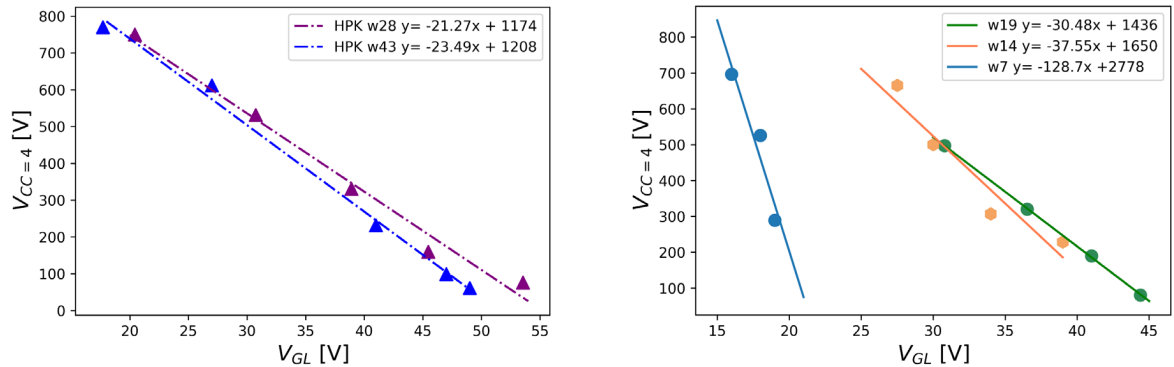

**Figure 4.** Correlation of VGL from 1/C2 measurement and VCC=4 from charge collection for HPK (Left) and FBK (Right) sensors.

## 7. Conclusions

Electrical tests done with a probe station and charge collection measurements were shown for both FBK and HPK sensor from the 2020 ATLAS and CMS prototype production. The doping concentration of the new production of HPK was optimized and the first split showed a better performance before irradiation than the old production while maintaining the same radiation hardness.
FBK sensors from the new production with deep gain layer and optimized doping and Carbon level showed exceptional radiation hardness allowing a gain up 20 and a time resolution of 40 ps at the HGTD maximum fluence of 2.5E15 $N_{eq}/cm^2$.